# Title: Regularity in forex returns during financial distress: Evidence from India

**Author**- R.P.Datta

Indian Institute of Foreign Trade, Kolkata Campus

1583 Madurdaha Chowbaga Road

Kolkata, PIN- 700107

West Bengal

India

Email- rpdatta@gmail.com

## Abstract

This paper uses the concepts of entropy to study the regularity/irregularity of the returns from the Indian Foreign exchange (forex) markets. The Approximate Entropy and Sample Entropy statistics which measure the level of repeatability in the data are used to quantify the randomness in the forex returns from the time period 2006 to 2021. The main objective of the research is to see how the randomness of the foreign exchange returns evolve over the given time period particularly during periods of high financial instability or turbulence in the global financial market. With this objective we look at 2 major financial upheavals, the subprime crisis also known as the Global Financial Crisis (GFC) during 2006-2007 and the recent Covid-19 pandemic during 2020-2021. Our empirical results overwhelmingly confirm our working hypothesis that regularity in the returns of the major Indian foreign exchange rates increases during times of financial crisis. This is evidenced by a decrease in the values of the sample entropy and approximate entropy before and after/during the financial crisis period for the majority of the exchange rates. Our empirical results also show that Sample Entropy is a better measure of regularity than Approximate Entropy for the Indian forex rates which is in agreement with the theoretical predictions.

# 1. Introduction

Shannon established the field of information theory in 1948 by incorporating the statistical physics notion of entropy into the channel communication process (Shannon 1948). Since then, other academics have made numerous modifications to Shannon's notion of entropy.
Kolmogorov developed the Kolmogorov entropy (KS entropy) (Kolmogorov 1985), which determines how quickly Shannon entropy changes in a system. Pincus (2008) offered approximation entropy (ApEn) as a substitute for short and noisy physiological time series in order to reduce the complexity of KS entropy computation, which necessitates complex mathematical reasoning.
The concept of regularity/irregularity in a sequential data set can be linked to the idea of randomness as follows. A completely random series may be understood as a completely irregular series and a completely regular series may be understood as a series with randomness zero. Further we can extend this logic to say a series that is completely random or completely irregular is also completely unpredictable while a series that has some regularity or patterns that repeat themselves are more predictable. The idea of using the concept of entropy in financial time series is to attempt to quantify the regularity/irregularity or in other words the extent of predictability/unpredictability in a financial time series. A time series that has repetitive patterns of changes is more predictable than a time series which do not have such patterns (Ho et al 1997). A financial time series that is more predictable offers a higher possibility of making profits than one which is less predictable or completely unpredictable that is random.

Approximate Entropy (ApEn) measures the possibility that similar patterns of observations will not be followed by more observations of the same kind. A series that has many repetitive patterns has a smaller Apen while a series that has less repetitive patterns has a higher value of ApEn. However, there are some shortcomings in the ApEn method that cause the findings to appear more regular than they actually are as pointed out by Bonal and Marshak (2019) and Richman and Moorman (2000).

Sample entropy (SampEn), first proposed by Richman and Moorman (2000), removes self-matching of the time series from the approximate entropy in order to obtain estimates of entropy that are more accurate. In this method two segments of data from a dataset of consecutive recorded points are considered to be matching if their distance falls below a value which is called the threshold value and which is fixed beforehand. The likelihood that two segments of a continuously measured dataset with one or more data points will still match if the two non-added portions are matched can be used to illustrate regularity. The negative logarithm of this probability is known as sample entropy, which measures irregularity (regularity) of the data set.

The approximate entropy and sample entropy both quantify the amount of regularity present in sequences of data, and they are independent of any underlying models. They have patterns as their foundation, and it is also possible to calculate both these statistics using only a limited number of data points.

This study uses the ApEn and SampEn statistics to evaluate and analyze the presence of repetitive patterns in the Indian Foreign Exchange rates vis a vis four foreign currencies namely the US Dollar (USD), the British Pound (GBP) , the Euro (EUR) and the Japanese Yen (Yen) with respect to the Indian Rupee (INR). The time period considered is January 2006 to December 2021 a total of 16 years. We calculate the logarithmic rate of return of USD/INR, GBP/INR, EUR/INR and JPY/INR during the entire period and analyze some basic features of the returns. Next we calculate the ApEn and SampEn values during two specific periods of global financial crisis namely the Global Financial Crisis (GFC) of 2007–2009 also known as the subprime crisis and the COVID-19 pandemic outbreak of 2020–2021 and compare their changes before and after the outbreak of the financial crisis.

In identifying the GFC's phases, scholars like Bartram and Bodnar (2009), Dooley and Hutchinson (20090, Olbrys and Majewska (2015), Claessens (2010, and the references therein, have expressed divergent views in the existing literature. The global financial crisis period for the U.S. and most of the European financial markets has been reported by Majewska (2017) and Olbrys (2022) as beginning in October 2007 and ending in February 2009. The outcomes are in line with the literature as seen in Bartram & Bodnar (2009) and Dooley & Hutchinson (2009).

The main working hypothesis that we propose in this paper is that when there is a sudden disruption in the financial markets, the entropy of the logarithmic returns in the major foreign exchange rates in India decreases. This indicates that during times of financial crisis the returns from the forex markets pertaining to India become more predictable and more regular. To test the working hypothesis empirically, the ApEn and SampEn values for the pre-crisis and post-crisis periods are computed.

The present research has two main contributions to the existing body of knowledge. Firstly, the empirical calculations and results support the working hypothesis. The values of both the SampEn and ApEn statistics for most of the exchange rates show a decline during the periods of financial instability thus finding no reason to reject our working hypothesis. Secondly, to the best of our knowledge there has been no empirical studies on the Indian foreign exchange rates using a model independent statistic like ApEn and SampEn. Though there have been some recent studies on the application of SampEn and ApEn in domains like stock markets, our findings in this regard can be considered as the first application of these statistics on Indian forex markets. The unique empirical results of this study, which have not yet been documented in the literature, are what give it its added worth.

Our results suggest that the financial time series studied here have repeated patterns i.e., exhibit some regularity, thus leading to higher predictability in the returns. These patterns show signs of increase during extreme event times. Thus, the possibility of predicting returns is supported by these findings, which are significant for academics and practitioners.

This is the first comparative study that, to the best of the present authors' knowledge, compares the returns from the 4 major foreign exchange rates in India as far as the presence of repeatable patterns in sequences of the return data as measured by the ApEn and SampEn statistics is concerned.

The remainder of this study is divided into the following sections. In the second section we present a short literature review of the existing body of knowledge. Section three describes the methodology behind the ApEn and SampEn algorithms. This is followed by section 4 which describes the source of our data and also some general findings pertaining to the data are reported. Section 5 discusses and compares the empirical results obtained for the four forex rates. We summarize our findings and point out some limitations and directions for future work in Section 6.

## 2. Literature Review

There have been many studies in recent years that have dealt with the application of entropy-based methods to study financial time series. (Wang and Shang (2018) apply multiscale entropy (MSE) to quantify the complexity of a financial time series. Chen et al (20200 use entropy plane analysis which combines the characteristics of complexity-entropy causality plane analysis and the theory of large deviations that depicts the spectral structure of the time series as a multifractal and apply it to both artificial and actual financial market data. Through the integration of time-varying effective transfer entropy (ETE) and several machine learning techniques, Kim et al (2020) , seek to forecast the direction of US stock prices. Namdari and Li (2019) examine the different uses of the general formulations of entropy-based uncertainty quantification and find that entropy metrics are effective predictors for stochastic processes with uncertainties.

Wang and Wang (2021) study the market performance of the S&P 500 Index, gold, Bitcoin, and the US Dollar Index in the context of the extreme COVID-19 pandemic event using a multiscale entropy-based method to measure the market efficiency at different time scales. In a paper Adam (2020) uses the daily stock market index for a number of African countries over a 10-year time period and the Rényi effective transfer entropy to look at the information flow from global economic policy uncertainty to emerging stock markets in Africa. Liu et al (2020) use entropy-based metrics to categorize various trading activities. By utilizing conditional block entropy, they are able to identify return-driven trading. Alvarez and Rodriguez (2021) suggested a method for evaluating the efficient market hypothesis (EMH) based on singular value decomposition (SVD) entropy.

Olbrys and Majewskawa (2020) in a comparative study evaluate the regularity and irregularity in stock market index time series in relation to their predictability between the COVID-19 pandemic (2020–2021) and for a two year period (2018–2019) before the pandemic. In another paper the same authors Olbrys and Majewskawa (2022) evaluate and compare changes in the regularity of 36 stock market indices in Europe and the United States during moments of significant financial turmoil. Ramirez et al (2021) define a nonparametric method of cross-sample entropy (CSE) estimation for long-memory and heteroskedastic models, considering a residual-based bootstrap-type estimator. Bonal (2021) used the Pincus index to compare the randomness of six most traded forex pairs with the intention of comparing their predictability. Their results show that the variance is simpler to predict for longer time frames than for shorter time horizons.

The above studies show that though there are some recent applications of the SampEn and ApEn statistics to analyze the stock returns of some countries there is not much research on the application of these statistics in the forex markets in general and India in particular. As mentioned in our conclusion, the nature of stock markets and forex market returns have some similarities and significant differences as well, see Jorion (1988), Giovannini & Jorion (1989) and Humala & Rodrigues (2013) for example. Therefore, our empirical study on the application of the SampEn and Apen statistics to analyze the Indian forex markets seem to be well justified.

## 3. Methodology

In this section we describe the methodology underlying the sample and estimate entropy calculations (ApEn and SampEn).

### 3.1 Approximate Entropy

Pincus (2009) proposed the ApEn algorithm to provide a novel statistic for experimental data series, as was mentioned in the introduction.The parameter ApEn evaluates the temporal and sequential regularity or irregularity of a series. Low ApEn values signify regularity, predictability, and recurrent patterns in the series. On the other hand, higher values of the approximate entropy imply that the data series is unpredictable and unstable and has few recurrent patterns Bonal (2019).

In this work, we put the ApEn algorithm's description from Bonal's paper into practice (Bonal 2019).

Consider the following time sequence v = {v(1), v(2), . . . , v( N )},
Where m is a positive integer , which satisfies 0≤ m ≤N, the length of the data subsets to be compared are denoted by N, and a real value r > 0, which indicates the tolerance level within which matches are admitted. For every computation, the variables N, m, and r must be fixed.

We then define the vectors $y_m$ (i ) = {v(i ), v(i + 1), . . . , v(i + m − 1)}
and $y_m$(j) = {v(j), v(j + 1), . . . , v(j + m -1)} . Then, the following formula is used to determine their Chebyshev distance::
d [ $y_m$ (i ), $y_m$ ( j)] = $\max_{k=1,2,...,m}$ (|v(i + k − 1) − v( j + k − 1)|).                                    (1)
The number of vectors $y_m$ ( j) within r of $y_m$ (i ) which allow self-counting is defined by Equation (2):

$$F_j^m (r) = \frac{1}{(N-M+1)} (number\ of\ j \leq (N - m + 1)\ such\ that\ d[y_m\ (i), y_m\ (j)] \leq r) \quad (2)$$

Given a predetermined threshold r, the numerator in Eq. 2 counts the number of vectors with consecutive lengths m that are equivalent to a given vector and enable self-counting.

The vector $y_m$(i) is referred to as the prototype when computing $F_j^m$ (r ), and a situation in which a vector $y_m$(j) lies within a distance r of it is referred to as a prototype match. $F_j^m$ (r ) is, in other words, the likelihood that any vector $y_m$(j) is within a distance r of $y_m$(i) (Richman & Moorman 2000).

The magnitude $F_j^m$ (r ) is calculated in the following step of the process using Eq. 3:

$$F^m(r) = \frac{1}{(N-M-1)} \sum_{i=1}^{N-m+1} log F_j^m (r) \quad (3)$$

Likewise if the dimension to be embedded is increased to m+1, the value of $F^{m+1}(r)$ is determined using Eq. 3. The time sequence v's estimated ApEn statistic value is given by:

$$ApEn(m, r, N)(v) = F^m(r) - F^{m+1}(r), \quad (4)$$

where m ≥ 1 and ApEn(0, r, N )(v) = −$F^1$(r).
The statistical estimator of the parameter ApEn(m, r) is given by the formula ApEn(m, r, N):
$$ApEn(m, r) = lim_{N \to \infty} [F^m(r) - F^{m+1}(r)]. \quad (5)$$

## 3.2 Sample Entropy

The calculations reported in this paper, implement the algorithm code for SampEn written in R which was reported in the paper by Bonal (2019), therefore a comparable nomenclature has been employed.
Consider the following time sequence again v = {v(1), v(2), . . . , v( N )}, a positive integer m, which satisfies 0≤ m ≤N, the length of the sequences to be compared denoted by N, and a real value r > 0, which indicates the tolerance level within which matches are admitted. For every computation, the variables N, m, and r must be fixed.

We then define the vectors $y_m(i) = \{v(i), v(i+1), \ldots, v(i+m-1)\}$
and $y_m(j) = \{v(j), v(j+1), \ldots, v(j+m-1)\}$. Then, the following formula is used to determine their Chebyshev distance:

$$d[y_m(i), y_m(j)] = \max_{k=1,2,\ldots,m}(|v(i+k-1) - v(j+k-1)|). \tag{6}$$

The next Equation (7) gives the maximum number of vectors $y_m(j)$ inside a radius r of $y_m(i)$ without allowing self-counting.

$$C_j^m(r) = \frac{1}{(N-M-1)} \sum_{j=1, j \neq i}^{N-M} \text{ (number of times that } d[y_m(i), y_m(j)] \leq r) \tag{7}$$

The total number of potential vectors $B^m(r)$ is then determined using Equation (8) which shows the empirical likelihood of two sequences matching for m points.

$$C_i^m(r) = \frac{1}{(N-M)} \sum_{i=1}^{N-M} C_i^m(r) \tag{8}$$

In a similar manner, the next Equation (9) states the maximum number of vectors $y_{m+1}(j)$ at a distance r of $Y_{m+1}(i)$ without permitting self-matching:

$$D_i^m(r) = \frac{1}{(N-M-1)} \sum_{j=1, j \neq i}^{N-M} \text{ (number of times that } d[y_{m+1}(i), y_{m+1}(j)] \leq r) \tag{9}$$

Next, we calculate the total number of matches $D^m(r)$ based on Equation (10). This denotes the empirical possibility that two sequences for m + 1 points (matches) are comparable.

$$D^m(r) = \frac{1}{(N-M-1)} \sum_{i=1,}^{N-M} D_i^m(r) \tag{10}$$

Since It is obvious that the number of matching vectors ($D^m(r)$) is always less than or equal to the number of likely vectors ($C_j^m(r)$) that match. Therefore, the ratio ($D^m(r)/C^m(r)$) <1 is a conditional probability (Shannon 1948).
The time sequence v's SampEn value is calculated in the final step, as shown below:

$$\text{SampEn}(m, r, N)(v) = -\log(D^m(r)/C^m(r)) \tag{11}$$

The SampEn(m, r, N) given by Equation (11) is the statistical estimator of the parameter SampEn(m, r):

$$\text{SampEn}(m, r) = \lim_{n \to \infty}[-\log(D^m(r)/C^m(r))] \tag{12}$$

The Sample Entropy is close to zero for regular, repeating data points because the quantity ($D^m(r)/C^m(r)$ in Equation (7) approaches one (Richman & Moorman 2004).

## 3.3 Comparing Sample Entropy and Approximate Entropy

ApEn and SampEn statistics have some significant distinctions, according to the research.
For example, the paper (Bonal 2019) provides a thorough explanation of these two approaches.

The equations for both statistics could be compared using the following methods after some algebraic manipulation:

$$ApEn(m, r, N) \approx \frac{-1}{(N-M)} \sum_{i=1}^{N-m} \log\left[\frac{\sum_{j=1}^{N-M}}{\sum_{j=1}^{N-M}} \left(\frac{\text{number of times that } d[y_{m+1}(i), y_{m+1}(j)] \leq r}{\text{number of times that } d[y_m(i), y_m(j)] \leq r}\right)\right], \tag{13}$$

$$SampEn(m,r,N) = -log\left[\frac{\sum_{i=1}^{N-M}\sum_{j=1,j\neq i}^{N-M}\left(\frac{number\ of\ times\ that\ d[y_{m+1}(i),y_{m+1}(j)] \leq r}{number\ of\ times\ that\ d[y_m(i),y_m(j)] \leq r}\right)}{\sum_{i=1}^{N-M}\sum_{j=1,j\neq i}^{N-M}}\right] \quad (14)$$

Where the same notation as in the previous subsections is used.

Based on the aforementioned equations Eq. 13–14, Delgado-Bonal and Marshak (2019)] underline some of the key points that differentiate between how ApEn and SampEn statistics are implemented.

a) The advantage of the SampEn technique given by Eq.12 is that it does not support self-counting (j ≠ I) while ApEn (Eq. 5) does allow self-counting. As this comparison yields no additional information, the SampEn algorithm avoids doing so.

b) The length of the series N enters the calculation of the ApEn statistic directly through the factor $\frac{1}{N-M}$ as shown in equation 13. This is perceived to be a drawback of the ApEn statistic as compared to the SampEn statistic.

c) SampEn is said to take into account the entire series because the total of all prototype vectors falls inside the logarithm, which implies that SampEn is already defined if a prototype finds a match.
For ApEn, on the other hand, each prototype must have a match.

Additionally, given the same data set, SampEn values are higher than ApEn values.

4. Data Description

In this paper we consider the exchange rate of 4 foreign currencies namely the US Dollar(USD), the British Pound(GBP), the Euro (EUR) and the Japanese Yen (JPY) with respect to the Indian rupee(INR).
The period of consideration is from 1st January 2006 to 31st December 2021. All the data used in this paper is obtained from the website investing.com. The equation below shows how the daily logarithmic returns of the foreign exchange rates with respect to the Indian Rupee are calculated:

$$R_t = ln\left(\frac{E_t}{E}\right) \quad (15)$$

Where $E_t$ is the value of the currency with respect to the Indian Rupee on day t.

The results of descriptive statistics of the log returns for the four exchange rates for the period are given in the Table below:

|  | USD | GBP | EUR | JPY |
|---|---|---|---|---|
| Mean |  |  |  |  |

|         |         | 0.00006 |         |         |
|---------|---------|---------|---------|---------|
|         | 0.00012 |         | 0.00010 | 0.00012 |
| Std dev | 0.00458 | 0.00654 | 0.00616 | 0.00767 |
| Kurtosis | 6.3472 | 6.7527 | 3.2391 | 6.1957 |
| Skewness | 0.1699 | -0.5072 | 0.07668 | 0.16146 |

Table1 - Descriptive statistics of the major foreign exchange rates

Some comments are in order on the observations in the Table above. We note that the mean of all the logarithmic returns for all the foreign exchange rates are very close to zero . Also the standard deviation of all the exchange rates are also very close to zero and there is not much variation between them for the time period considered. From Table 1 we observe that the distributions of all the returns are slightly positively skewed except for GBP which is negatively skewed. The values of the kurtosis from Table 1 indicate that the returns from USD, GBP and JPY are strongly leptokurtic (value of skewness above 6) while the kurtosis value for EUR is close to 3 i.e. the normal distribution. These observations are generally in consonance with the characteristics of returns from financial time series as observed in the literature (Tsay 2010).

We have also conducted the Augmented Dickey Fuller (ADF) test on the logarithmic returns for all the foreign exchange rates and found that the series of logarithmic returns is stationary. The results of the ADF test are not reported here for sake of brevity but are available to the interested reader on request.

The plot of the log returns for the various currencies are given below in Figures 1 to 4..

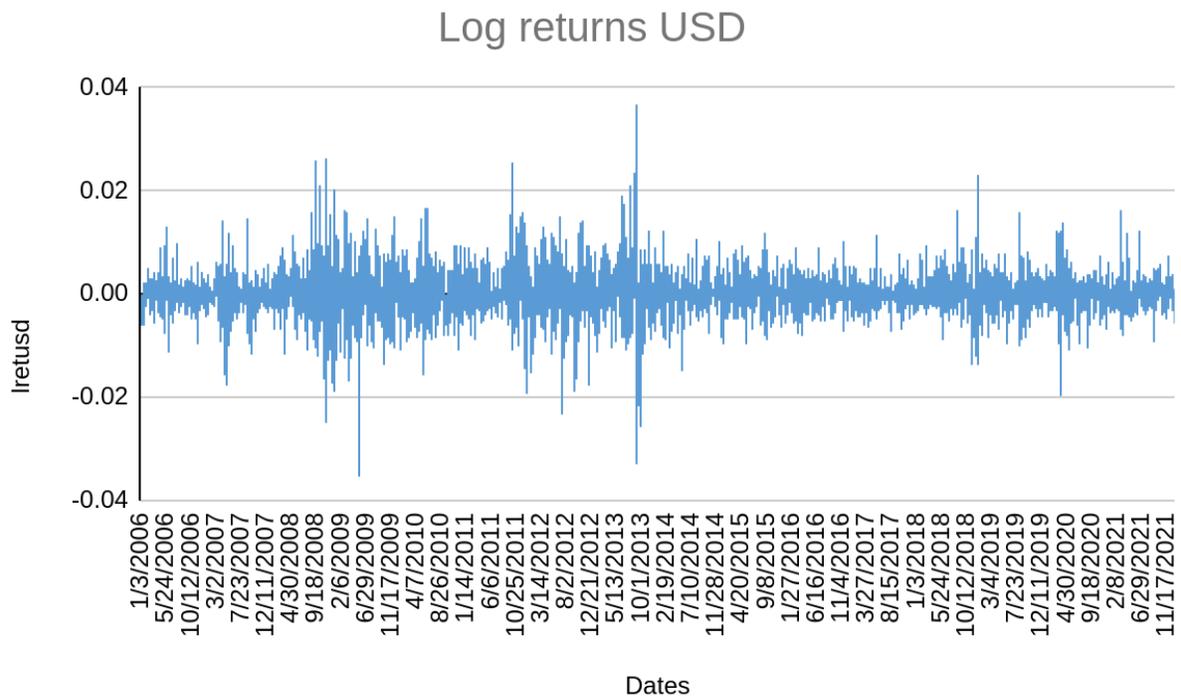

Figure-1 - Logarithmic returns of USD/INR

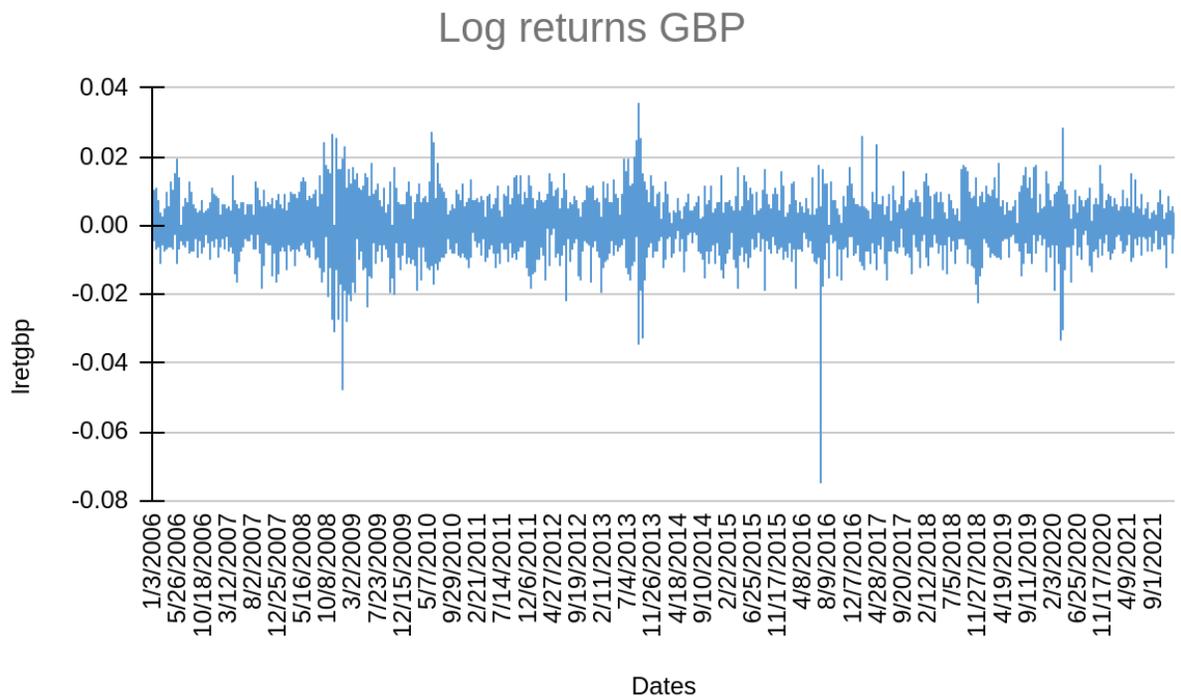

Figure -2- Logarithmic returns of GBP/INR

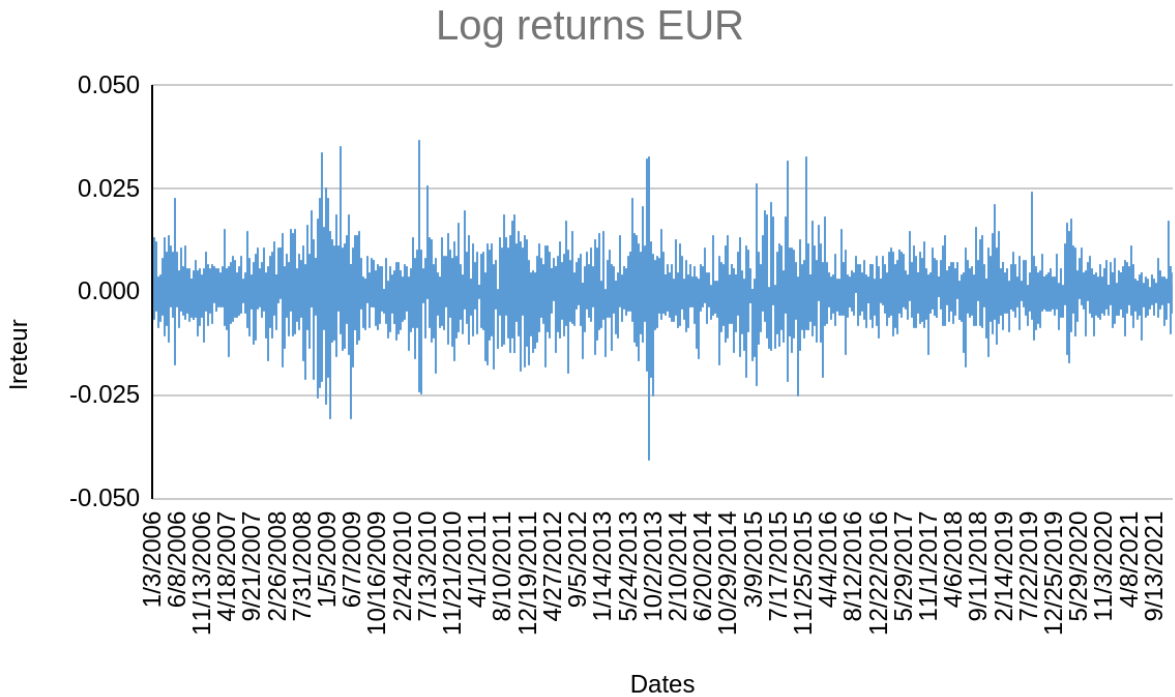

Figure-3- Logarithmic returns of EUR/INR

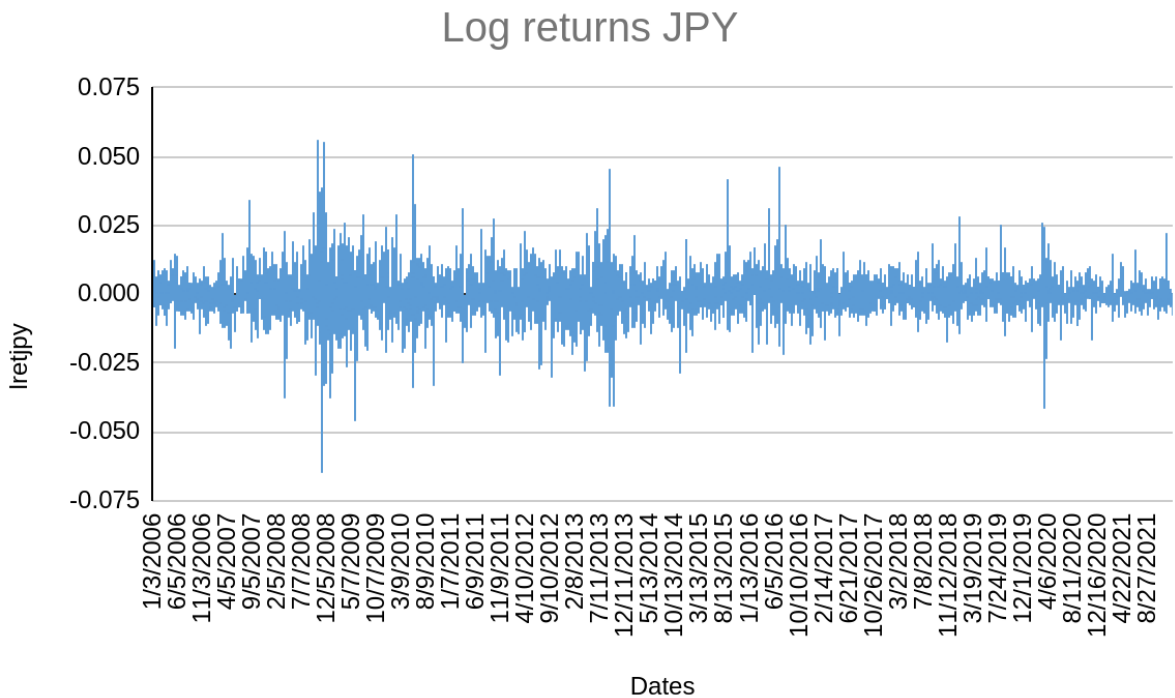

Figure-4-Logarithmic returns of JPY/INR

A few qualitative observations may be made from the Figures 1 to 4. The logarithmic returns of all the major foreign exchange currencies with respect to the Indian rupee exhibit clusters of high volatility. Most of the exchange rates show some commonality like there is a period of high volatility from the middle of 2008 to the middle of 2009, during the second half of the year 2013 and during the first half of 2020. However the time period during which other volatility clusters are observed are different for different foreign exchange rates. The reasons behind these may be attributed to the fact that the economic events that affect the exchange rates do not influence all the exchange rates in the same manner. The Indian economy is an emerging economy and the extent of its integration with respect to the global economy has changed during the time period under consideration. The Indian Rupee is not yet fully convertible. Though it has been fully convertible in the current account since 1994 it is only partially convertible in the capital account. The major part of Indian foreign currency transactions are done through the US Dollar. Also the Reserve Bank of India intervenes frequently in the forex market in line with various policy initiatives. The effect of an economic event on the USD/INR exchange rate may not be the same on the GBP/INR, EUR/INR and JPY/INR exchange rates.

## 5. Results

We now present the results from our empirical calculations to look at the change in the approximate Entropy and Sample entropy during 2 major periods of financial stress. The time periods considered in our research are given below:

i) From May 2006 to September 2007, before the global financial crisis (Total length 17 months).

ii) From October 2007 to February 2009, following the subprime crisis (Total length 17 months).

iii) From January 2018 to December 2019, before the COVID-19 pandemic phase (total length 24 months)

iv) From January 2020 to December 2021, following the COVID-19 pandemic phase (total length 24 months);

The time periods were chosen with two specific periods of economic uncertainty in mind namely the General Financial Crisis (GFC) also known as the subprime crisis and the Covid 19 pandemic. Both these events severely affected the global financial markets. The periods of economic uncertainty were chosen based on the references by Bertram et al (2009), Dooley et al (2009), Majewska (2017), Olbrys (2022 c) and WHO (2022) .

There are 2 parameters that need to be input for calculating the SampEn and ApEn values, namely m which is the length of the consecutive data points to be considered for matching purposes and r the acceptable level till which a match is considered. The choice of m and r is more of a practitioner's choice. Olbrys and Majewska (2022) state that m should be 1 or 2, while Bonal and Marshak (2019) state that m values of 2 or 3 are more typical. More number of matching prototypes are found for m equal to 1, while a m value of 2 can give more information on the nature of the evolving sequences. In this work we choose m=1. As for the value of r, we follow Richman and Moorman (2004) where r is taken to be 0.2 times the standard deviation of the data points under consideration i.e. r = 0.2*sd , where sd is the standard deviation.

The following tables present the results from our calculations:

Table 2 - Sample Entropy results for the 4 exchange rates before and after General Financial Crisis

|       | SampEn pregfc | SampEn postgfc | Change   |
| ----- | ------------- | -------------- | -------- |
| Forex |               |                |          |
| USD   | 1.7176        | 1.6331         | Decrease |
| GBP   | 2.3103        | 1.9018         | Decrease |
| EUR   | 1.9469        | 1.9915         | Increase |
| JPY   | 1.9231        | 1.8960         | Decrease |

Table 3 -Sample Entropy results for the 4 exchange rates before and after Covid 19 pandemic

|       | SampEn precovid | SampEn postcovid | Change   |
| ----- | --------------- | ---------------- | -------- |
| Forex |                 |                  |          |
| USD   | 1.9107          | 1.9613           | Increase |
| GBP   | 1.9954          | 1.8078           | Decrease |
| EUR   | 2.1666          | 1.8611           | Decrease |
| JPY   | 2.0318          | 1.8332           | Decrease |

Table 4- Approximate Entropy results for the 4 exchange rates before and after General Financial Crisis

|       | ApEn pregfc | ApenEn postgfc | Change   |
| ----- | ----------- | -------------- | -------- |
| Forex |             |                |          |
| USD   | 1.2031      | 1.1657         | Decrease |
| GBP   | 1.2530      | 1.1719         | Decrease |
| EUR   | 1.1517      | 1.1634         | Increase |
| JPY   | 1.2179      | 1.2374         | Increase |

Table 5 - Approximate Entropy results for the 4 exchange rates before and after Covid 19 pandemic

|  | ApEn precovid | ApEn postcovid | Change |
|---|---|---|---|
| Forex |  |  |  |
| USD | 1.3 | 1.3565 | Increase |
| GBP | 1.3375 | 1.2793 | Decrease |
| EUR | 1.3966 | 1.3543 | Decrease |
| JPY | 1.4289 | 1.4262 | Decrease |

We make the following observations:

1) From Table 2 we see that the Sample Entropy for the USD, GBP and JPY have decreased from the pre GFC period to the post GFC period while for EUR the Sample Entropy has increased.
2) From Table 3 we see that the Sample Entropy for the GBP, EUR and JPY have all decreased during the phase before the COVID-19 pandemic to the post COVID-19 phase while for the USD the Sample Entropy has increased.
3) From Table 4 we see that the Approximate Entropy for the EUR and JPY decreases while the Approximate Entropy for the USD and GBP increase from the pre GFC to the post GFC period.
4) From Table 5 we see that the Approximate Entropy for the GBP,EUR and JPY decreases while the Approximate Entropy for the USD increases from before the onset of Covid-19 to the post Covid 19 time phase.
5) The Sample Entropy method shows that 6 (3 pre GFC to post GFC and 3 pre Covid-19 to post Covid-19) out of 8 possible values (4 pre GFC to post GFC and 4 before Covid-19 to after Covid-19) for the 4 exchange rates show a decrease in their values during the periods of financial uncertainty/turbulence.
6) The Approximate Entropy method shows that 5(2 pre GFC to post GFC and 3 before Covid-19 to after Covid-19) out of 8 possible values (4 pre GFC to post GFC and 4 before Covid-19 to after Covid-19) for the 4 exchange rates show a decrease in their values during the periods of financial uncertainty/turbulence.

It is usually accepted in the literature that Sample Entropy(SampEn) is a better measure of randomness or predictability than Approximate Entropy (ApEn) since the latter allows self counting as explained in the methodology section before. Our empirical findings confirm this as we see that the calculated values of the Sample Entropy before and after the periods of severe financial instability agree more with our working hypothesis i.e. periods of high instability in the financial markets lead to more predictive behavior in the values of financial assets. The empirical findings from ApEn, though still in favor of the working hypothesis, are more diverse.

Another point particularly of relevance while interpreting the SampEn results (which is a better representation of the order of regularity in the data series as mentioned before in this paper) is the value of the input parameter m. We have pointed out before that m was chosen to be 2 for our calculations and the reasons for this choice have also been mentioned.
For sake of numerical experimentation we repeated our calculations with m=1 and found the results to be similar. In fact for the SampEn case if m is chosen to be 1 instead of 2, we find the value of SampEn for all the foreign exchange rate returns, decreases both after the Global Financial crisis during 2006-2007 and the Covid-19 period. The results were actually in better agreement with our working hypothesis that

SampEn values should decrease during periods of financial instabilty. All the 8 cases ( 4 pre GFC to post GFC, and 4 pre Covid to post Covid) for the USD, GBP, EUR and JPY show a reduction in the SampEn values for m=1. The numerical results for m=1 for SampEn are not shown in this paper to avoid repetition and for the sake of brevity.

Having said that a few facts pertaining to the Indian Foreign Exchange markets should be taken into consideration as well. As mentioned before in this paper, the Indian Rupee is still not fully convertible in the capital account till date. The currencies that are chosen here on the other hand are all fully convertible globally. The integration of Indian financial markets has evolved over the time period of this study from less integrated to more integrated. During times of extreme financial distress the Reserve Bank of India has responded with policy interventions that are not in consonance with the countries whose currencies are taken here for exchange rate calculations. Also the effects of the GFC and the Covid-19 pandemic on the foreign exchange rates are not simultaneous in nature.

In a previous study Olbrys (2022 a ), it has been shown that considering the stock market indices of several countries, the working hypothesis that stock markets seem to become less random and hence more regular or predictable during periods of high financial distress, can not be rejected. Our empirical results on the foreign exchange markets in India are in agreement with these findings.

# 6. Conclusion

To conclude, our empirical calculations of the Sample Entropy (SampEn) and Approximate Entropy (ApEn) of the exchange rates of 4 currencies namely the USD, GBP, EUR and the JPYwith respect to the Indian Rupee before and after 2 periods of high financial distress in the global financial markets seem to suggest that the working hypothesis that randomness or predictability increase during periods of financial crisis can not be rejected. In fact, the majority of the results on the SampEn and ApEn calculations suggest that indeed the values of the SampEn and ApEn decreased during a financial crisis from before the financial crisis. This is understandable because during periods of financial instability traders operating in the foreign exchange markets tend to become risk averse.  In other words in some sense the forex markets in India reacted in a predictable manner during periods of extreme financial distress.

Recent studies by Olbrys (2022 a and b ) applying the SampEn and ApEn statistics conclude that there is an increase in the regularity of stock market returns during times of financial turbulence. The forex markets in general have a different structure than stock markets as suggested by Jorion (1988) and Giovannini & Jorion (1989). Humala & Rodrigues (2013) have found different non-normalities in the forex and stock markets along with fat tails, excess kurtosis, return clustering, and unconditional time-varying moments. This study has found that cycles of volatility in the currency and stock markets are linked to frequent occurrences of macroeconomic financial instability in the case of Peru. Our empirical results suggest that the forex markets in India also exhibit similar behavior i.e increase in regularity during periods of financial crisis as observed in the case of stock markets in the studies mentioned above.

Some limitations of our study concern the time periods chosen for demarcating the periods of financial turbulence. It has been argued that during the GFC period the Indian forex market was comparatively insulated from the global markets due to stringent regulations imposed by the Reserve Bank of India. Therefore, the exact period of financial turbulence that affected the Indian economy may not be the same as the time period chosen in this paper. The same is true for the Covid-19 pandemic period as well. Particularly the Indian government imposed one of the most stringent lockdowns in the world in late March 2020. Also, the opening up of the Indian economy post lockdown was in phases and the effect on factors that affect the foreign exchange rates may not exactly coincide with the period considered in this paper.